\documentclass[12pt]{article}

\usepackage[utf8]{inputenc}
\usepackage{amsmath}
\usepackage{amssymb}
\usepackage{hyperref}
\usepackage{physics}
\usepackage{graphicx}
\usepackage{geometry}
\usepackage{epstopdf}
\usepackage{authblk}

\title{Chiral phase transition in soft-wall AdS/QCD with scalar-dilaton coupling}
\author{Sean P. Bartz\thanks{Email: sean.bartz@indstate.edu}}
\author{Robert C. Meadows}
\author{Glenn Brock}
\affil{Dept.~of Chemistry and Physics, Indiana State University, Terre Haute, IN 47809}

\begin{document}

\maketitle

\begin{abstract}
The chiral phase boundary of nuclear matter is expected to have a critical point where the rapid crossover of lattice methods at zero chemical potential becomes a first-order phase transition.
 Phenomenological models based on the AdS/CFT correspondence, known as  AdS/ QCD, have succeeded in capturing many features of nuclear matter, with recent progress in producing the expected critical point. We study a model that produces a critical point in the chiral phase diagram by introducing a coupling between the scalar chiral field and the dilaton. We examine the effect of the scalar-dilaton coupling on the critical point. We also study the zero-temperature chiral dynamics, which must allow for spontaneous chiral symmetry breaking in the limit of zero quark mass. 
 %When the scalar-dilaton coupling is too small, the correct chiral dynamics are not produced. Conversely, when the coupling is too large, the critical point disappears. 
We find that when the scalar-dilaton coupling is large enough to ensure correct zero-temperature chiral dynamics, a critical point is present only if the quark mass is greater than 12.8 MeV. 
\end{abstract}

\section{Introduction}

The exploration of the phase structure of quantum chromodynamics (QCD) at extreme temperature and density is an important project for both theory and experiment \cite{Elfner_2023_heavy_ion_review,AN_2022_BEST_framework, Pandav2022}. 
Lattice QCD finds a crossover phase transition at zero quark chemical potential \cite{Aoki2006,AARTS_2023_Lattice_phase_transitions_review}, while  other models find a first-order phase transition at high chemical potential \cite{FUKUSHIMA201399}. In combination, these models suggest the existence of a critical point, but its exact location in the phase diagram remains an open question \cite{Bellwied2016, Gunkel2021_CEP_theory}. 

Experimentally, the search for the critical point requires a reduction in center of mass energy \cite{Adamczyk2014}, motivating the recently-completed Beam Energy Scan at the Relativistic Heavy Ion Collider \cite{BES2010}, as well as current and future fixed-target experiments \cite{Meehan2017,Ablyazimov2017,FAIR2022}.
From the theoretical perspective, lattice methods suffer from a well-known sign problem at finite chemical potential \cite{Philipsen2013}. Extrapolation techniques allow lattice analysis up to a baryon chemical potential $ \approx 300$ MeV, with no evidence of a critical point \cite{Borsanyi_2020_lattice_finite_mu}.

The AdS/CFT correspondence \cite{Maldacena1998TheSupergravity, Witten1998Anti-deTheories, Witten1998AntiHolography} has emerged as a powerful tool to study various aspects of QCD, including the phase diagram \cite{Gubser2008, Gubser2008a}. The soft-wall AdS/QCD model, which uses a background dilaton field to encode confinement, has been extensively used to analyze hadron spectra \cite{karch-katz-son-adsqcd,Karch2006LinearAdS/QCD, Erlich2005QCDHadrons} and the QCD phase diagram \cite{ Rougemont2024}.
While there has been success in finding a critical point in the deconfinement phase transition using holographic techniques \cite{DeWolfe2011, DeWolfe2011a,Li2017a, Critelli2017,Rougemont2018, Grefa2021,Cai2022, hippert2023_bayesian_holographic_CEP}, producing a critical point in the chiral phase transition has been more elusive.

In this work, we consider a modified  soft-wall AdS/QCD model with a coupling between the scalar and dilaton fields. Prior work  has shown that the introduction of such a coupling can improve the resulting meson spectra and introduce a critical point in the chiral phase diagram \cite{Fang2019}. We focus on the effect of scalar-dilaton coupling on the critical point in the QCD phase diagram and the zero-temperature chiral dynamics.

\section{Soft-wall model with scalar-dilaton coupling} 
We use an anti-de Sitter black hole metric
\begin{equation}
    ds^2=\frac{L^2}{z^2}\left(-f(z)dt^2+dx_i^2+\frac{dz^2}{f(z)}\right),\label{metricGeneric}
\end{equation}
with the AdS curvature $L=1$ throughout the rest of this work.
Following  established procedure \cite{Chamblin1999ChargedHolography,Park2010DissociationMedium,Colangelo2011HolographyDiagram},  we model finite temperature and chemical potential with a charged black hole described by the  5D AdS--Reissner-Nordstr{\"o}m blackness function
\begin{equation}
    f(z)=1-(1+Q^2)\left(\frac{z}{z_h}\right)^4+Q^2\left(\frac{z}{z_h}\right)^6, \label{metricMu}
\end{equation}
where $z_h$ is the black hole horizon, and $Q$ is related to the black hole charge $q$ by $Q=qz_h^3$. The quark chemical potential and temperature are determined by the charge and horizon position
\begin{eqnarray}
\mu&=&\kappa \frac{Q}{z_h}, \\
T&=&\frac{1}{\pi z_h}\left(1-\frac{Q^2}{2}\right),
\end{eqnarray} 
where  $0<Q^2<2$ and $\kappa=1$ \cite{Colangelo2012TemperatureStudy}.
Note that $\mu$ is the quark chemical potential, with a value one third of the baryon chemical potential.  These relationships are invertible for $z_h$ and $q$.

In the soft-wall model of AdS/QCD,  confinement is introduced via a dilaton field, which is quadratic in the IR limit. In this work, we use a quadratic dilaton
\begin{equation}
\Phi(z) = \mu_g^2 z^2 \label{eq:dilaton}
\end{equation}
where $\mu_g=440$ MeV sets the confinement scale. 
The relevant matter fields are described by the  action
\begin{equation}
\mathcal{S}=\frac{1}{2k} \int d^5x \sqrt{-g} e^{-\Phi(z)}  \left\{\mathrm{Tr}\left[|DX|^2+V_m(X,\Phi)\right]+\gamma \textrm{Re}\left[\textrm{det}(X) \right] \right\},\label{fullaction}
\end{equation}
where $X$ contains the scalar and pseudoscalar meson fields. The t'Hooft determinant term  provides mixing between light and heavy flavors, and with $\gamma<0$ it produces a first-order phase transition in the chiral limit \cite{Chelabi2016ChiralAdS/QCD}. We omit the vector and axial-vector meson fields to focus on the chiral dynamics. 

A quartic term in the scalar potential is required for spontaneous chiral symmetry breaking \cite{gherghetta-kelley}.
Including the coupling between the dilaton and scalar field, the potential is
\begin{equation}
V_m(X,\Phi) = m_5^2 |X|^2  + \lambda_1 \Phi |X|^2 + 4\lambda_4 |X|^4,
\end{equation}
The AdS/CFT dictionary sets the mass of the chiral field $m_5^2=-3$ \cite{Witten1998AntiHolography}. 
 Following \cite{Fang2018}, we take $\lambda_4=4.2$ and $\gamma=-22.6$.

\begin{figure}[tb]
    \centering
    \includegraphics[width=0.7\textwidth]{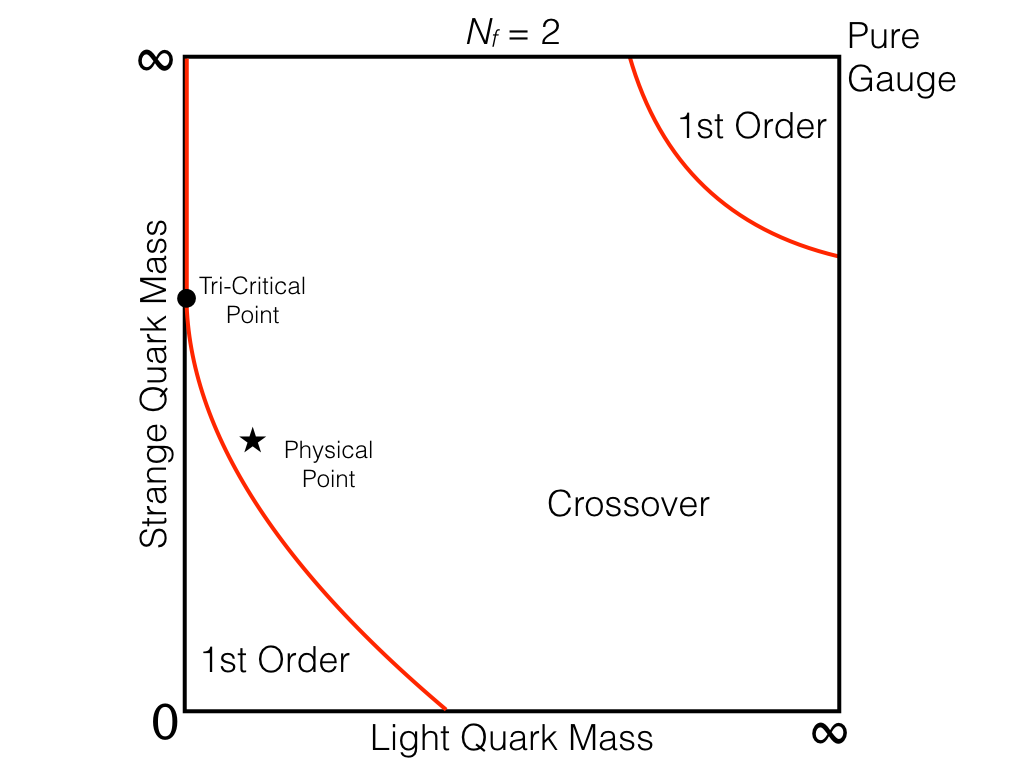}
    \caption{A sketch of the Columbia plot, which shows the expected order of the chiral phase transition as a function of light and strange quark masses \cite{columbiaplot_brown, deForcrand:2006pv}.}
    \label{fig:columbia}
\end{figure}

The scalar-dilaton coupling term gives the chiral field an effective mass that runs with energy scale
\begin{equation}
    m_5^2 \rightarrow -3 + \lambda_1 \mu_g^2 z^2. \label{eq:running_mass}
\end{equation}
This running mass has been used to obtain the correct mass splitting between excited states of meson chiral partners \cite{Fang2016ChiralAdS/QCD}, to reproduce the Columbia plot (Figure \ref{fig:columbia}) at zero chemical potential \cite{Fang2018},  to obtain the correct chiral transition behavior \cite{Fang_2020}, and to produce a critical point in the chiral phase diagram \cite{Fang2019}.
It is worth noting that models with a modified dilaton profile but without the running mass (\ref{eq:running_mass}) achieve the first three of these goals \cite{Chelabi2016ChiralAdS/QCD, Bartz:2016ufc} but do not produce the critical point \cite{Bartz2018_2plus1}. 

The scalar  field has a $z$-dependent vacuum expectation value (VEV) that describes the chiral symmetry breaking of the model. In a three flavor model, the VEV becomes
\begin{equation}
\langle X \rangle = \frac{1}{\sqrt{2}}\begin{pmatrix}
{\chi_u(z)} & 0 & 0 \\
0 & {\chi_d(z)} & 0 \\
0 & 0 & {\chi_s(z)}
\end{pmatrix}.
\end{equation}
In this work, we will focus on the flavor-symmetric case $\chi_u=\chi_d=\chi_s$. 
Varying \ref{fullaction} with respect to $\chi$  yields the following equation of motion,
\begin{equation}
\chi''-\left( \frac{3}{z} +\Phi'-\frac{f ' }{ f } \right)\chi' -\frac{1}{z^2f} \left[ (-3-\lambda_1 \Phi)\chi + 4\lambda_4 \chi^3 +3 \lambda_3 \chi^2 \right]=0, \label{eq:chiEOM}
\end{equation}
%The value $\lambda_1=7.438$ corresponds to $\mu_c=1180$ MeV, as used in paper on running mass by Fang, et al.
where $\gamma \rightarrow 6\sqrt{2}\lambda_3$ is defined for convenient notation.
As the chiral field is the source of the $\bar{q}q$ operator, the AdS/CFT dictionary identifies its coefficients at the UV boundary with the sources of chiral symmetry breaking,
\begin{equation}
    \chi(z\rightarrow 0) \sim m_q \zeta z + \frac{\sigma}{\zeta}z^3,
\end{equation}
where $\zeta=\sqrt{N_c}/(2\pi)$ \cite{Cherman:2008eh}, the quark mass $m_q$ is the source of explicit chiral symmetry breaking, and the chiral condensate $\sigma$ is the source of spontaneous chiral symmetry breaking.

\section{Numerical procedure} \label{sec:numerical}
Finding the chiral condensate requires solving (\ref{eq:chiEOM}) numerically and using the AdS/CFT dictionary to relate the solution for $\chi(z)$ to the parameters $m_q, \, \sigma$. The presence of a singular point at $z=z_h$ presents a challenge to this procedure. 
A commonly-used numerical method begins with a UV approximation for the chiral field and integrates toward the horizon \cite{Li2013a,Bartz:2016ufc,Fang2016ChiralAdS/QCD}. While this method works well near the chiral transition temperature, it is less reliable at low temperatures.  
Instead, we use a method that starts with the asymptotic solution at the black hole horizon $z_h$ and integrates toward the UV boundary \cite{BallonBayona2021}. A comparison between these two methods is discussed in Appendix \ref{sec:numerical}.

The near-horizon solution is approximated by the Taylor series 
\begin{equation}
    \chi(u \rightarrow 1) = d_0 +d_1 (1-u)+d_2 (1-u)^2 + \ldots
\end{equation}
where $u=z/z_h$ and the higher-order coefficients are solved by substitution into (\ref{eq:chiEOM}). The result is
\begin{eqnarray}
    d_1 &=& \frac{d_0}{2 \left( Q^2 -2 \right)}\left( 3 +  \lambda_1  z_h^2\mu_g^2 - 3 d_0 \lambda_3 - 4 d_0^2 \lambda_4 \right)
 \\
 d_2 &=& \frac{1}{{16 \left( Q^2 -2 \right)^2}} \Big\{ 6 d_1 (-6 + Q^2 + Q^4) + 4 d_0^3 (14 - 13 Q^2) \lambda_4 \nonumber \\
 &+& d_0^2 \left[(42 - 39 Q^2) \lambda_3   - 24 d_1 \left( Q^2 -2 \right) \lambda_4 \right] - 2 d_1 \left( Q^2 -2 \right)  ( 4 Q^2 -8 - \lambda_1)z_h^2 \mu_g^2\nonumber \\
 &+& 3 d_0 \left[-14 + 13 Q^2 + 8 d_1 \lambda_3 - 4 d_1 Q^2 \lambda_3 + \lambda_1( 3 Q^2-2 ) z_h^2  \mu_g^2\right] \Big\}
\end{eqnarray}

For each value of $T,\mu$, we vary $d_0$ and compare the numerical solution to the UV expansion of the chiral field
\begin{eqnarray}
\chi(u\rightarrow 0)& \approx & m_q \zeta z_h u -3m_q^2 \zeta^2 \lambda_3  z_h^2 u^2 +\frac{\sigma}{\zeta} z_h^3 u^3 \nonumber \\
&+& \frac{1}{4}\left[ m_q^3\zeta^3 (\lambda_4 - 36 \lambda_3^2) +2m_q \zeta \mu_g^2(\lambda_1-2) \right] z_h^3 u^3 \log \left(z_h u\right) + \dots \label{eq:chiral_UV}
\end{eqnarray}
The terms of order $u$ and $u^3$ have their coefficients defined by the AdS/CFT dictionary and the other coefficients are found by solving (\ref{eq:chiEOM}) order-by-order.

% \begin{eqnarray}
%     m_q  &=& \frac{1 - \sqrt{1 - 3 \lambda_3 (3 \chi(u_i) - u_i\chi'(u_i) )}}{3 \lambda_3 \zeta u_i z_h}\\
%     \sigma &=&  \zeta \frac{ 1 - 3\lambda_3(2 \chi(u_i)  +   u_i\chi'(u_i) ) - \sqrt{1 - 3 \lambda_3 (3 \chi(u_i) - u_i\chi'(u_i) )}}{3 u_i^3 z_h^3 \lambda_3} 
% \end{eqnarray}
We evaluate the numerical solution and its derivative at a small value  $u_i\approx 10^{-3}$ and calculate the coefficients by comparing to the UV expansion (\ref{eq:chiral_UV}).
Keeping terms up to order $u^3$, the relationships are analytically solvable,
\begin{eqnarray}
    m_q  &=& \frac{1 - \sqrt{1 - 3 \lambda_3 (3 \chi_{UV} - u_i\chi'_{UV} )}}{3 \lambda_3 \zeta u_i z_h},\\
    \sigma &=&  \zeta \left(\frac{ 1 - 3\lambda_3(2 \chi_{UV}  +   u_i\chi'_{UV} ) - \sqrt{1 - 3 \lambda_3 (3 \chi_{UV} - u_i\chi'_{UV} )}}{3 u_i^3 z_h^3 \lambda_3} \right),
\end{eqnarray}
where $\chi_{UV} = \chi(u_i)$ and $\chi'_{UV}=\chi'(u_i)$. The quadratic relationship has another set of solutions, which produce unphysical values of $m_q, \sigma < 0$. %Keeping additional terms from (\ref{eq:chiral_UV}) requires a numerical solution that does not improve the accuracy of the solutions but adds significant computation time.
In the two-flavor case, $\lambda_3=0$ and the above relationships cannot be used. Instead we find 
\begin{eqnarray}
    m_q &=& \frac{3\chi_{UV}-u_i\chi'_{UV}}{2\zeta z_h u_i}, \\
    \sigma &=& \zeta \frac{u_i \chi'_{UV}-\chi_{UV}}{2 z_h^3 u_i^3}.  
\end{eqnarray}

% In the two-flavor case, $m_q$ and $d_0$ have a one-to-one relationship, so the required value of $m_q$ is found via root finding with the numerical shooting method, and the corresponding value of $\sigma$ is then calculated.
% In the three-flavor case, each $m_q$ value is associated with up to three $d_0$ values, corresponding to up to three $\sigma$ values. Thus, it is necessary to generate a plot of $m_q$ vs. $d_0$ and read off the $d_0$ values that produce the input $m_q$ value. The corresponding $\sigma$ values are then calculated in the same way. \todo{Is this necessary?}

\section{Results}

\begin{figure}[htb]
    \centering
    \includegraphics[width=\textwidth]{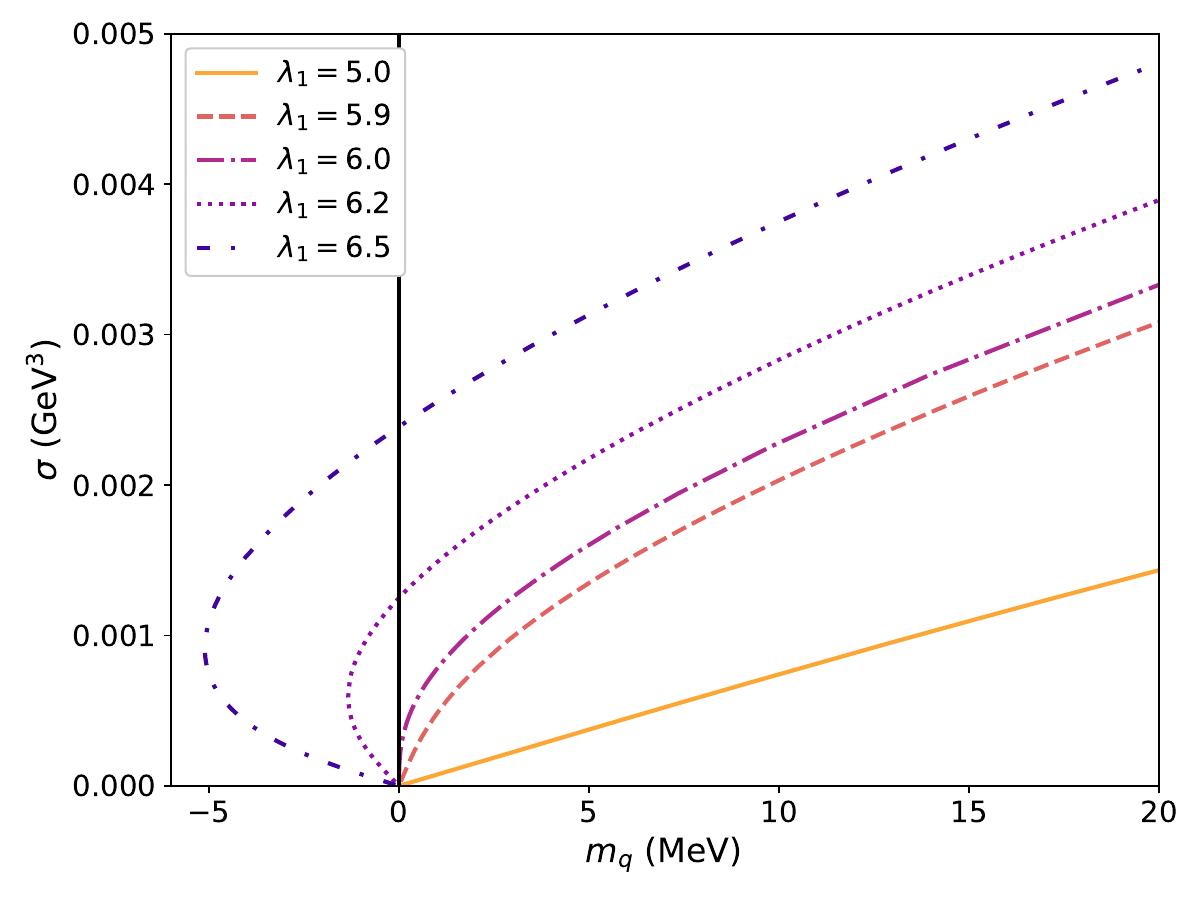}
    \caption{The relationship between $\sigma$ and $m_q$ for a variety of values of the scalar-dilaton coupling $\lambda_1$ with two flavors at zero temperature and chemical potential. For $\lambda_1 \ge 6.0$, the chiral condensate is present even in the chiral limit $m_q=0$.}
    \label{fig:sigma_vs_mq_2flavor}
\end{figure}

In this section, we find the dependence of the chiral condensate $\sigma$ on quark mass $m_q$ for various values of the scalar-dilaton coupling $\lambda_1$. We show that there is a minimum value of $\lambda_1$ that allows for spontaneous chiral symmetry breaking in the chiral limit $m_q\rightarrow 0$.
We show how the (pseudo-) critical temperature is found and how crossover and first-order transitions are distinguished. Finally, we show how the location of the critical point is affected by the value of $\lambda_1$.

\subsection{Spontaneous chiral symmetry breaking} \label{sec:chi_SB}
% \begin{figure}
%     \centering
%     \begin{minipage}{0.49\textwidth}
%         \centering
%         \includegraphics[width=\linewidth]{sigma_vs_mq_T0_2flavor_subset-eps-converted-to.pdf}
%         \end{minipage}\hfill
%     \begin{minipage}{0.49\textwidth}
%         \centering
%         \includegraphics[width=\linewidth]{sigma_vs_mq_T0_3flavor_subset-eps-converted-to.pdf}
        
%     \end{minipage}
%     \caption{The relationship between the chiral condensate $\sigma$ and quark mass $m_q$ for a variety of values of the scalar-dilaton coupling $\lambda_1$ for two flavors (left) and three flavors (right). In the three-flavor case, when the plot is multi-valued, the lower value is thermodynamically favored. In both cases, $sigma \rightarrow 0$ in the chiral limit when $\lambda_1<6$. \label{fig:sigma_vs_mq}}
% \end{figure}

Separate sources of spontaneous and explicit chiral symmetry breaking are required in the theory. The original soft-wall model did not achieve this, finding  $\sigma \sim m_q$ instead \cite{karch-katz-son-adsqcd}. Including a quartic term in the scalar potential allows these quantities to be independent,  crucially maintaining spontaneous chiral symmetry breaking  in the chiral limit $m_q=0$ \cite{gherghetta-kelley}.

The relationship between $m_q$ and $\sigma$ in this model depends on the strength of the scalar-dilaton coupling $\lambda_1$. We check this relationship in both the 2-flavor ($\lambda_3=0$) and 3-flavor ($\lambda_3 \neq 0$) cases. 
In the two-flavor case, the relationship between $\sigma$ and $m_q$ is one-to-one for quark mass $m_q \ge 0$. For small values of the scalar-dilaton coupling, the spontaneous chiral symmetry breaking vanishes as $m_q \rightarrow 0$, but at higher values of $\lambda_1 \geq 6.0$, $\sigma$ is non-zero in the chiral limit, as seen in Figure \ref{fig:sigma_vs_mq_2flavor}.

\begin{figure}[tb]
    \centering
    \includegraphics[width=\textwidth]{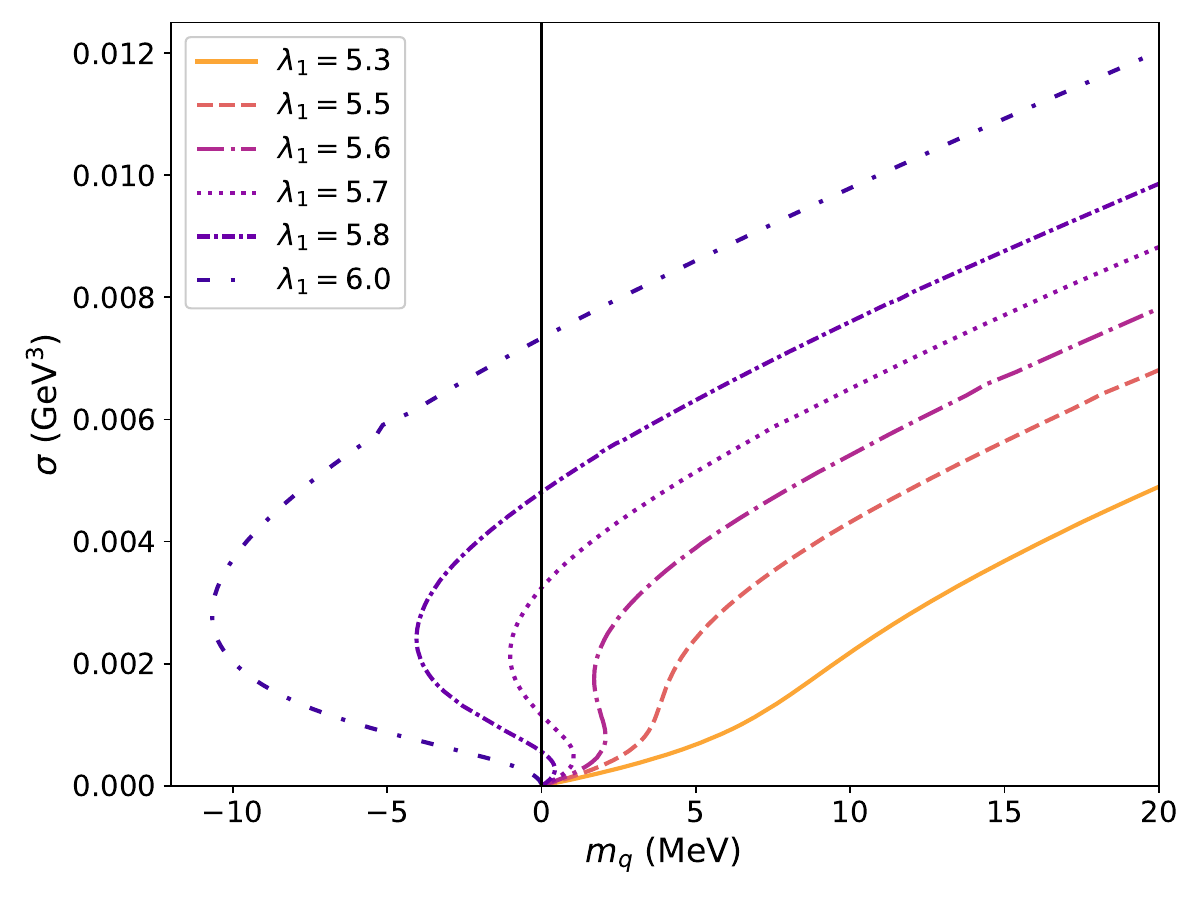}
    \caption{The relationship between $\sigma$ and $m_q$ for a variety of values of the scalar-dilaton coupling $\lambda_1$ with three flavors. When $\sigma$ is multi-valued, the lowest value is thermodynamically favored. For $\lambda_1 \ge 6.0$, the chiral condensate is present even in the chiral limit $m_q=0$.}
    \label{fig:sigma_vs_mq_3flavor}
\end{figure}

In the three-flavor case, we find the same requirement $\lambda_1\ge 6.0$ for $\sigma$ to remain finite as the quark mass goes to zero. In Figure \ref{fig:sigma_vs_mq_3flavor}, it is evident that $\sigma$ becomes multi-valued for intermediate values of $\lambda_1$. In these cases, the smaller value of $\sigma$ is  thermodynamically favored. This means that in the case of e.g. $\lambda_1=5.7$ it appears that there are finite solutions of $\sigma$ at zero quark mass, but these solutions are unphysical, and the lower branch of the graph shows $\sigma \sim m_q$ at small quark mass for these values of $\lambda_1$.

% \begin{figure}[tb]
%     \centering
%     \includegraphics[width= \textwidth]{Temp_dependent_sigma_vs_mq_lambda1=5.6-eps-converted-to.pdf}
%     \caption{The relationship between $\sigma$ and $m_q$ is temperature-dependent. A representative selection of temperatures is shown here in the three-flavor case for $\lambda_1=5.6$ and $\mu=0$.} 
%     \label{fig:temp_dependent_sigma_mq}
% \end{figure}

% \begin{figure}[tb]
%     \centering
%     \includegraphics[width=\textwidth]{mu_dependent_sigma_vs_mq_lambda1=5.6-eps-converted-to.pdf}
%     \caption{The relationship between $\sigma$ and $m_q$ is dependent on the chemical potential. A representative selection of $\mu$ values is shown here in the three-flavor case for $\lambda_1=5.6$ and $T=0$.}
%     \label{fig:mu_dependent_sigma_mq}
% \end{figure}

\subsection{Chiral phase transition}

\begin{figure}[tb]
  \centering
  \begin{minipage}[b]{0.49\textwidth}
    \includegraphics[width=\textwidth]{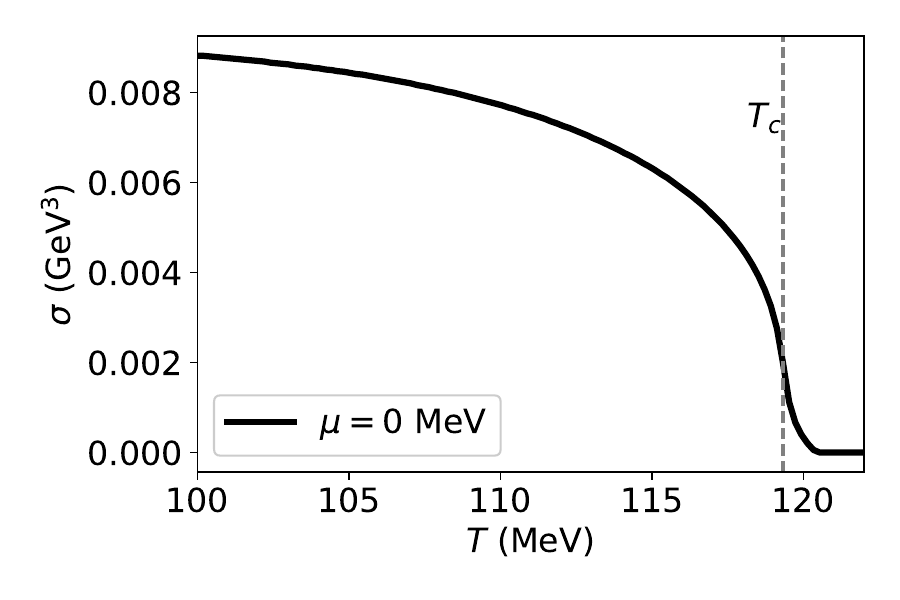}
  \end{minipage}
  \begin{minipage}[b]{0.49\textwidth}
    \includegraphics[width=\textwidth]{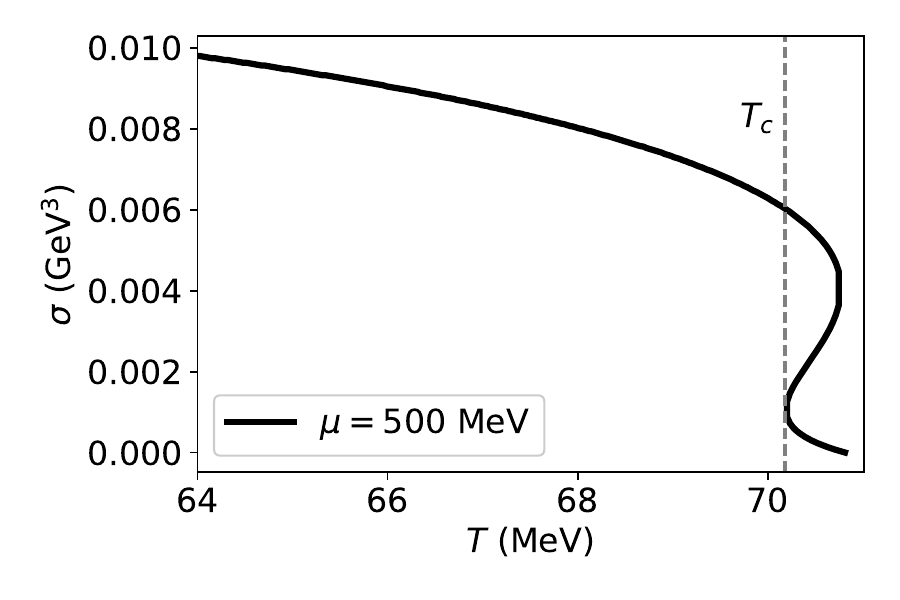}
  \end{minipage}
  \caption{These plots illustrate the difference between a crossover phase transition and a first-order phase transition. At zero chemical potential (left), the transition is smooth, but at higher $\mu$ (right), the value of $\sigma$ becomes multi-valued, indicating a first-order phase transition. Both plots use $m_q=15$ MeV and $\lambda_1=6$. \label{fig:sigma_T}} 
\end{figure}

% \begin{figure}[tb]
%     \centering
%     \includegraphics[width=0.75\textwidth]{sigma_vs_T_lambda1_compare4_mq=1_mu=0_plasma_extra_sigfigs-eps-converted-to.pdf}
%     \caption{ Decreasing the scalar-dilaton coupling parameter  $\lambda_1$ increases the prominence of the unphysical ``bump" before the chiral phase transition. When $\lambda_1$ is small enough, chiral symmetry is restored at both low and high temperatures, which is not physically accurate. The above data is for $m_q=1$ MeV and zero chemical potential.}
%     \label{fig:chiral_transition_lambda1_effects}
% \end{figure}

For a given value of the chemical potential $\mu$ and quark mass $m_q$, we find the values of $\sigma$ for a range of temperatures.
The order of the chiral phase transition  is determined by the way in which $\sigma$ transitions to a smaller value. Smooth transitions are considered crossover, while a first-order phase transition is characterized by the chiral condensate becoming multi-valued, as illustrated in Figure \ref{fig:sigma_T}.

\begin{figure}[tb]
    \centering
    \includegraphics[width=\textwidth]{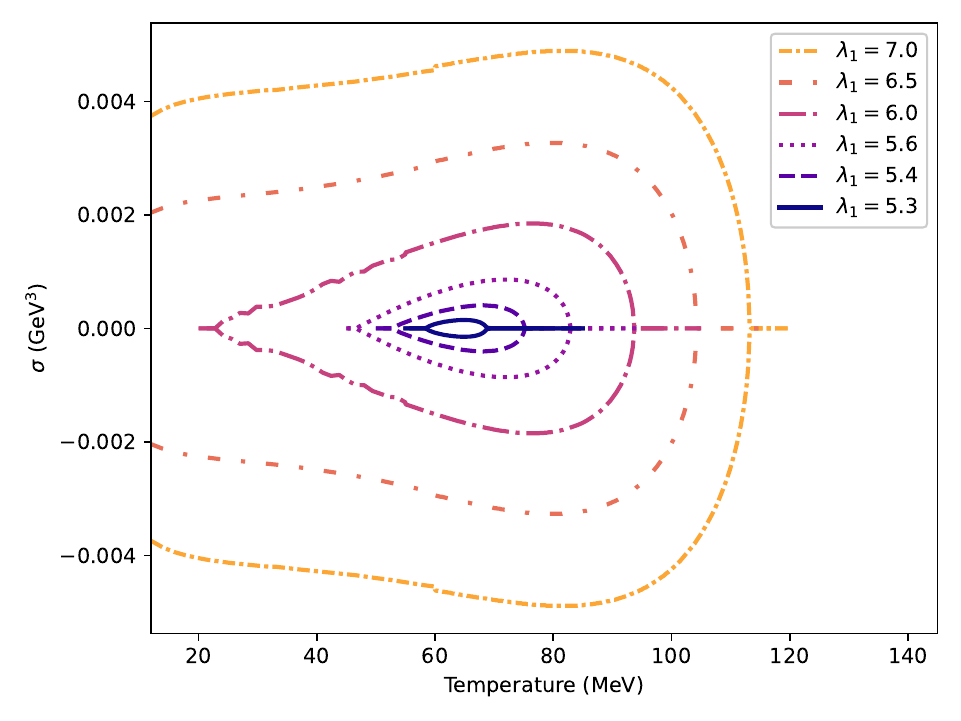}
    \caption{The two-flavor results in the chiral limit for chiral condensate $\sigma$ as a function of temperature at zero chemical potential. For scalar-dilaton coupling $\lambda_1 \le 6$, $\sigma$ vanishes at low temperatures.}
    \label{fig:chiral2flavor}
\end{figure}

In crossover transitions, the pseudo-critical temperature is the temperature where the chiral susceptibility $|d\sigma/dT|$ is maximized. For first-order transitions, the critical temperature occurs at the lowest temperature where $\sigma$ is multi-valued.

\begin{figure}[tb]
    \centering
    \includegraphics[width=\textwidth]{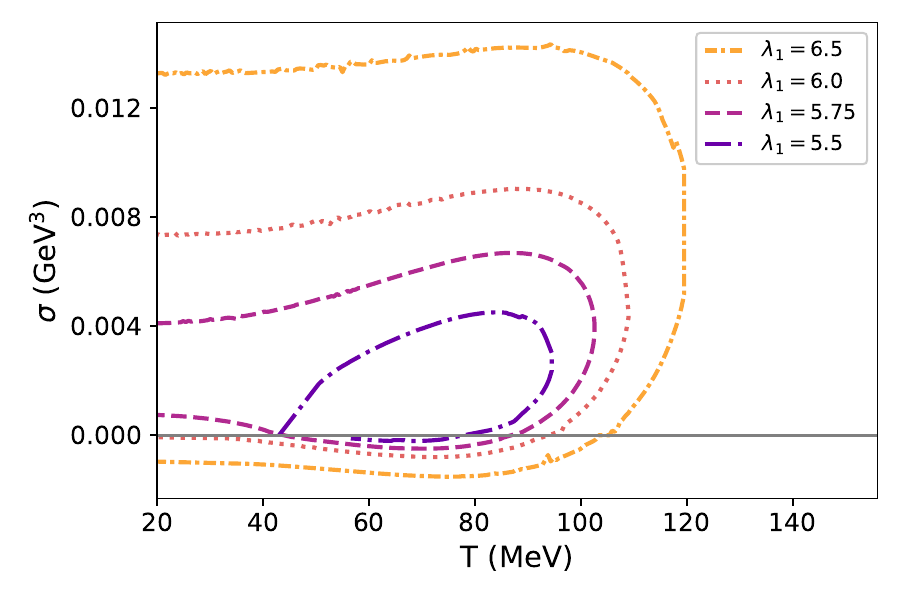}
    \caption{The three-flavor results in the chiral limit for chiral condensate $\sigma$ as a function of temperature at zero chemical potential. For scalar-dilaton coupling $\lambda_1 \le 5.5$, $\sigma$ vanishes at low temperatures.}
    \label{fig:chiral3flavor}
\end{figure}

The values of $\lambda_1$ that produce unphysical chiral dynamics at zero temperature also show unphysical behavior in the chiral phase transition. The chiral condensate is plotted as a function of temperature at $\mu=0$ in the chiral limit for 2 flavors (Figure \ref{fig:chiral2flavor}) and 3 flavors (Figure \ref{fig:chiral3flavor}). Note that the two-flavor case allows  negative values of $\sigma \leftrightarrow -\sigma$ as solutions. This symmetry is broken in the three-flavor case, and also when $m_q>0$, although negative solutions are still present \cite{Chelabi2016ChiralAdS/QCD}. These non-physical solutions are ignored in the rest of the analysis. 

The chiral phase transition has the expected low $T$ behavior in the chiral limit. As this parameter is decreased, the ``bump" below the critical temperature becomes more pronounced. When $\lambda_1$ is below a certain value, the chiral condensate disappears at low temperature. This unphysical result is further evidence for a minimum value for the scalar-dilaton coupling in this model.

\subsection{Phase diagram and critical point}

The chiral phase diagram is produced by plotting the (pseudo-) critical temperature as a function of the chemical potential. A critical end point is found for combinations of $m_q$ and $\lambda_1$ that produce a crossover phase transition at zero chemical potential. At sufficiently large $\mu$, the phase transition becomes first order.

\begin{figure}[tb]
   \centering
 \includegraphics[width=0.75\textwidth]{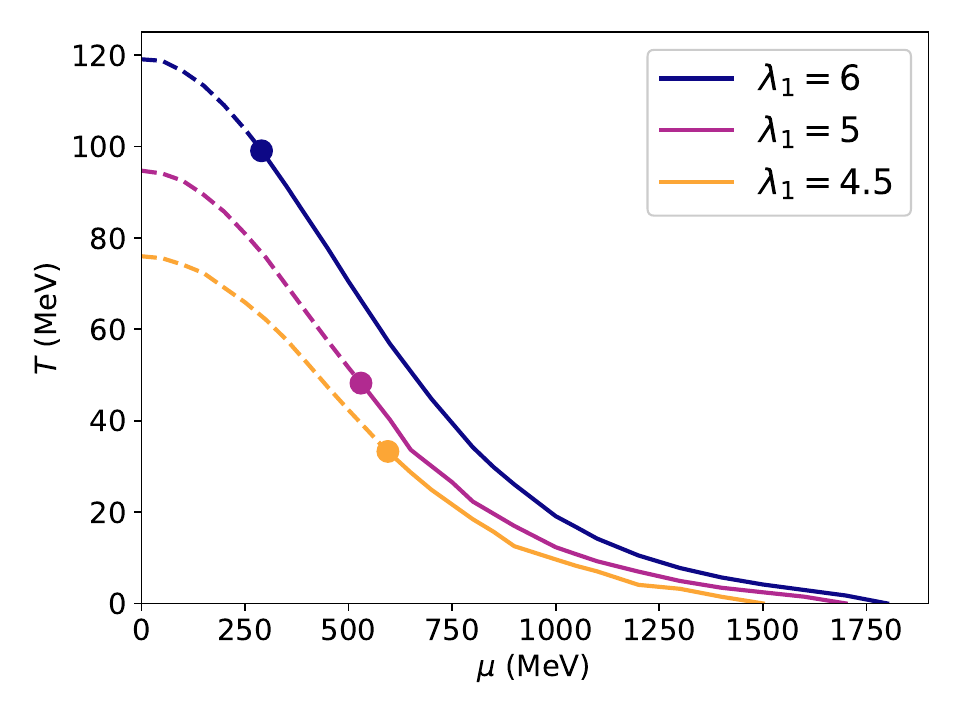}
 \caption{The phase diagram is shown for $m_q=15$ MeV for various values of the scalar-dilaton coupling parameter $\lambda_1$. Dashed lines indicate a  crossover and solid lines indicate a first-order phase transition. The critical points are indicated by a large dot. Increasing $\lambda_1$ moves the critical point to smaller chemical potential values while also increasing the crossover temperature at $\mu=0$.   }
   \label{fig:PhaseDiagram_mq_15} 
\end{figure}

To examine the effect of scalar-dilaton coupling on the location of the critical point, we show in Figure \ref{fig:PhaseDiagram_mq_15} phase boundaries for a sample quark mass $m_q=15$ for varying $\lambda_1$ with three symmetric quark flavors. The critical point occurs at smaller chemical potential as $\lambda_1$ is increased. At the same time, the (pseudo-) critical temperature is increased at all values of $\mu.$  

\begin{figure}[tb]
    \centering
    \includegraphics[width=\textwidth]{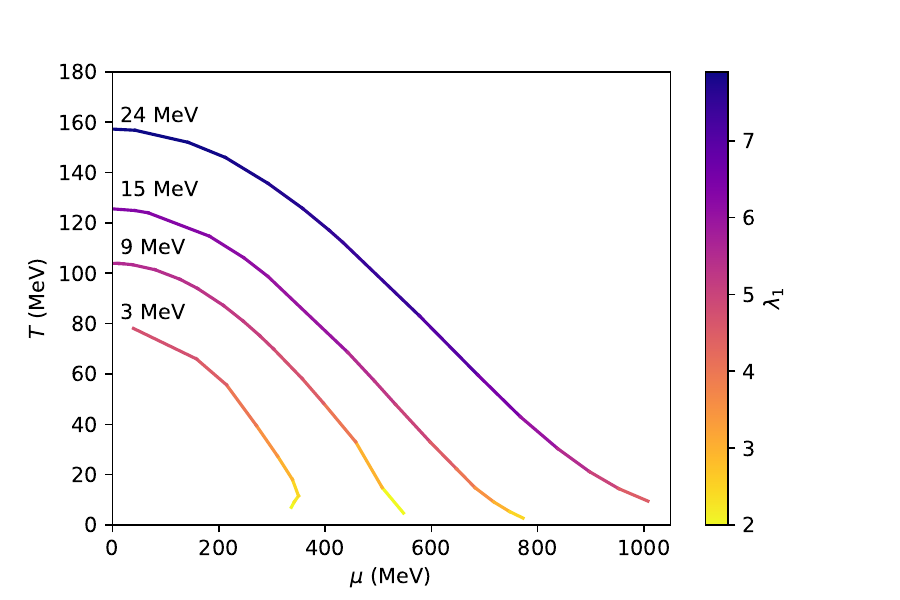}
    \caption{The locations of the critical point for different values of the quark mass and scalar-dilaton coupling $\lambda_1$ are shown. The curves are labeled by the value of the quark mass.}
    \label{fig:critical_points}
\end{figure}

When the scalar-dilaton coupling is sufficiently large, the phase transition is first order at $\mu=0$, and the critical point vanishes. This is seen in Figure \ref{fig:critical_points}, where the location of the critical point is plotted for several values of the quark mass. When the quark mass is large, a critical point can still be found when $\lambda_1 \ge 6.0$, as required for the proper chiral dynamics detailed in Section \ref{sec:chi_SB}. At low values of the quark mass, obtaining a critical point requires $\lambda_1 < 6.0$. We find that the minimum quark mass with a critical point in the phase diagram when $\lambda_1 = 6.0$ is $m_q=12.8$ MeV.
In the chiral limit, we find that the chiral phase transition is always first-order, regardless of the value of $\lambda_1$, and no critical point is present.

\section{Discussion}

In this work, we used a soft-wall holographic QCD model with a scalar-dilaton coupling to study the chiral phase transition at finite temperature and density. Our analysis shows that a coupling $\lambda_1 \geq 6.0$ is necessary for achieving the correct chiral dynamics with either two or three symmetric quark flavors, in line with the  requirement that  holographic  models  mirror the chiral symmetry breaking mechanism of QCD \cite{Karch2006LinearAdS/QCD, Erlich2005QCDHadrons}.
Furthermore, the presence of a critical point in this model's chiral phase diagram contributes to the ongoing effort to comprehend the phase structure of QCD, a topic of considerable theoretical and experimental interest \cite{RAJAGOPAL1999150, Laermann:2003cv}. %The existence and location of the critical point have profound implications for the understanding of the phase transitions in QCD, potentially guiding future experimental searches in heavy-ion collisions \cite{Bazavov2012ChiralTransition}.

Previous soft-wall AdS/QCD models that achieved correct chiral dynamics at zero chemical potential by using a UV-modified dilaton \cite{Chelabi2016ChiralAdS/QCD, Bartz2018_2plus1} could be extended to include a scalar-dilaton coupling term. It may be interesting to examine whether this will circumvent the problems shown in the current work at small values of the scalar-dilaton coupling.

Looking ahead, we will allow the strange quark mass to differ from the light quark masses and explore the 2+1 flavor results at a range of scalar-dilaton coupling. These results will be compared to the Columbia plot shown in Figure \ref{fig:columbia}. 
Another goal is to combine the analysis of the chiral transition with the deconfinement phase transition. Previous work combining a scalar chiral field with the dynamical Einstein-Maxwell-Dilaton model have shown some promise in the case of two quark flavors \cite{YANG2022}. Including the scalar-dilaton mixing term produces the crossover chiral transition that is expected for two quark flavors with $m_q>0$  \cite{Fang2023}.  
Considering these extensions will allow a more thorough exploration of all aspects of the holographic QCD phase diagram.

\section*{Acknowledgments}
Work on this project was partially funded by a grant from the University Research Committee at Indiana State University. Further support came from Indiana State's Summer Undergraduate Research Experience. SB thanks Alfonso Ballon-Bayona and Diego Rodrigues for useful discussion about numerical methods.

\bibliographystyle{utphys}
\bibliography{thebib.bib}

\appendix

\section{Comparison of numerical methods} \label{sec:comparison}

\begin{figure}[htb]
    \centering
    \includegraphics[width=0.7\textwidth]{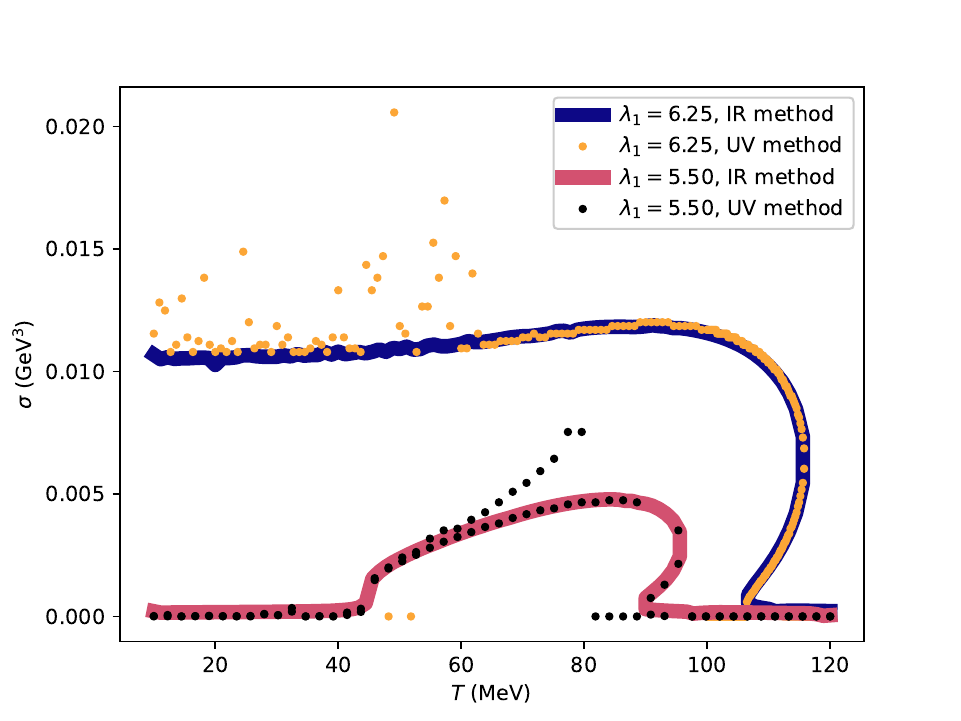}
    \caption{This comparison of the numerical methods shows agreement between them, but more spurious points when integrating from the UV. The data shown is for two representative values of the scalar-dilaton coupling $\lambda_1$ with $m_q=1$ MeV and $\mu=0$.}
    \label{fig:method_compare}
\end{figure}
Another commonly-used numerical method for solving the chiral equation of motion (\ref{eq:chiEOM})  uses the UV expansion of (\ref{eq:chiral_UV}) as one boundary condition and the regularity of the chiral field at the horizon $z=z_h$ is used as the other.
Regularity is difficult to check numerically, so a ``test" function is defined that includes all the potentially singular parts of the equation of motion,
\begin{equation}
   -z^2 \frac{f'(z)}{f(z)}\chi'(z) +\frac{1}{f(z)}\left[ (-3-\lambda_1 \Phi)\chi + 4\lambda_4 \chi^3 +3 \lambda_3 \chi^2 \right].
\end{equation}
This collection of terms must be zero at the horizon, otherwise there will be a divergence as $f\rightarrow 0$. Ensuring the test function is zero becomes the second boundary condition.

The shooting method \cite{NumericalRecipes3} is implemented with the required quark mass given as an input parameter and $\sigma$  varied until the boundary condition is met near the black hole horizon.
This  method is used to find the allowed values of $\sigma$ for a given $T, \mu$ and, $m_q$. %In the three-flavor case, there may be up to three $\sigma$ values at a given temperature, signifying a first-order phase transition.

The limitations of this method are revealed at low temperature $T \ll T_c$. At low temperature, $z_h \sim T^{-1}$ becomes large, and numerical instabilities in the numerical solution to (\ref{eq:chiEOM}) make it difficult to determine the correct value of $\sigma$. This has not been a problem in previous publications that focus on the chiral phase transition. Typically, the value of $\sigma$ approaches a constant value as temperature is decreased below the transition temperature.  However, this is not always the case in this model, where $\sigma =0$ at low temperatures for some values of $\lambda_1$, as seen in Figure \ref{fig:method_compare}.

The method of integrating from the UV boundary is trustworthy near the transition temperature.
However, in this work we are also interested in the low temperature chiral dynamics, particularly in ensuring separate sources of explicit and spontaneous chiral symmetry breaking,  as discussed in Section \ref{sec:chi_SB}. 
By starting near the singular point and integrating away from it, the method of Section \ref{sec:numerical} is more numerically stable at lower temperatures.

\end{document}